\documentclass[aps,prb,showpacs,twocolumn]{revtex4}
\usepackage{amssymb}
\usepackage{amsmath}
\usepackage{graphicx}
\usepackage{subfigure}

\begin{document}

\title{Confinement-enhanced spin relaxation for electron ensembles in large quantum dots}
\author{E. J. Koop}
\author{B. J. van Wees}
\author{C. H. van der Wal}
\affiliation{Physics of Nanodevices Group, Zernike Institute for
Advanced Materials, University of Groningen, Nijenborgh 4, NL-9747AG
Groningen, The Netherlands}
\date{\today}

\begin{abstract}
    We present a numerical study of spin relaxation in a
    semiclassical electron ensemble in a large ballistic
    quantum dot. The dot is defined in a GaAs/AlGaAs
    heterojunction system with a two-dimensional electron gas, and
    relaxation occurs due to Dresselhaus and Rashba spin
    orbit interaction. We find that confinement in a micronscale dot
    can result in strongly enhanced relaxation with respect to a free
    two-dimensional electron ensemble,
    contrary to the established result that strong confinement or
    frequent momentum scattering
    reduces relaxation. This effect occurs when the size of
    the system is on the order of the spin precession length, but smaller than the mean free path.
\end{abstract}


% 73.20.Fz Weak or Anderson localization
% 72.25.Dc Spin polarized transport in semiconductors
% 72.25.Hg Electrical injection of spin polarized carriers
% 72.25.Rb Spin relaxation and scattering
% 73.23.Ad Ballistic transport
% 73.23.-b Electronic transport in mesoscopic systems
% 73.23.Hk Coulomb blockade; single-electron tunneling
% 73.40.-c Electronic transport in interface structures
% 73.50.Gr Charge carriers: generation, recombination, lifetime,
%                           trapping, mean free paths
% 73.63.Kv Quantum dots

% See http://publish.aps.org/PACS/

\maketitle

% $\renewcommand{\baselinestretch}{2}

%%%%%%%%%%%%%%%%%%%%%%%%%%%%%%%%%%%%%%%%%%%%%%%%%%%%%%%%%%%%%%%%%%%%%%%%%
%%% INTRODUCTION %%%%%%%%%%%%%%%%%%%%%%%%%%%%%%%%%%%%%%%%%%%%%%%%%%%%%%%%
%%%%%%%%%%%%%%%%%%%%%%%%%%%%%%%%%%%%%%%%%%%%%%%%%%%%%%%%%%%%%%%%%%%%%%%%%

Due to spin-orbit interaction (SOI), the state of electron
spins is influenced by electron transport in electronic
devices. This has been recognized as a source for dephasing
and relaxation for spins \cite{8Silsbee2004,8Fabian2007},
as well as a means for controlled spin manipulation
\cite{8Datta1990} in research that aims at developing
spintronic devices
\cite{8Fabian2007,8Zutic2004,8Awschalom2007}. In this
article we present a numerical study of spin relaxation in
an electron ensemble that is scattering inside a
micronscale device structure. We are interested in the case
where devices are made of clean semiconductor
heterostructures and studied at low temperatures. For our
studies we assume realistic material parameters for a
system with a two-dimensional electron gas (2DEG) at a
GaAs/AlGaAs heterojunction. For a free 2DEG in these
materials, a well established result is that the average
spin orientation of an ensemble decays due to precession in
spin-orbit fields. For moderate electron mobilities, this
so-called D'yakonov-Perel' (DP) mechanism for spin
relaxation \cite{8Dyakonov1971,8Dyakonov1986} has the
property that the spin relaxation time $T_{1}$ increases
when the mobility (and thereby the time scale $\tau_{s}$
for elastic momentum scattering) decreases, as $T_{1}
\propto \tau_{s}^{-1}$.

%%%%%%%%%%%%%%%%%%%%%%%%%%%%%%%%%%%%%%%%%%%%%%%%%%%%%%%%%%%%%%%%%%%%%%%%%
%%% FIGURE 1 %%%%%%%%%%%%%%%%%%%%%%%%%%%%%%%%%%%%%%%%%%%%%%%%%%%%%%%%%%%%
%%%%%%%%%%%%%%%%%%%%%%%%%%%%%%%%%%%%%%%%%%%%%%%%%%%%%%%%%%%%%%%%%%%%%%%%%
\begin{figure}
\includegraphics[width=0.67\columnwidth]{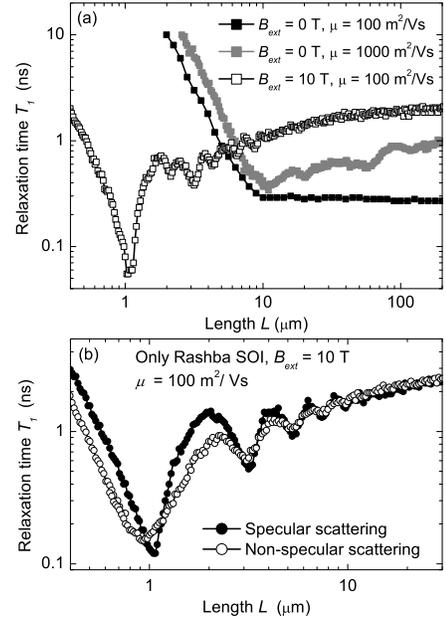}
\caption{(a) Spin ensemble relaxation in a quantum dot
system of size $L$. The relaxation time $T_1$ is calculated
as a function the size $L$ for a square system. For
mobility $\mu = 100~{\rm m^2/Vs}$ we plot $T_1$ for zero
external magnetic field (filled black symbols) and for
$B_{ext} = 10~{\rm T}$ (open symbols). The gray symbols
show $T_1$ for zero external magnetic field and $\mu =
1000~{\rm m^2/Vs}$. (b) Relaxation time as a function of
$L$ for a system with only Rashba SOI. Calculations for
specular and non-specular reflections give qualitatively
the same results, but the magnitude of the resonant
structure in the traces of $T_1$ as a function of $L$ is
larger in the case of specular reflections.}
\label{8Fig:Trelax_vs_length}
\end{figure}
%%%%%%%%%%%%%%%%%%%%%%%%%%%%%%%%%%%%%%%%%%%%%%%%%%%%%%%%%%%%%%%%%%%%%%%%%

%%%%%%%%%%%%%%%%%%%%%%%%%%%%%%%%%%%%%%%%%%%%%%%%%%%%%%%%%%%%%%%%%%%%%%%%%
%%% FIGURE 2 %%%%%%%%%%%%%%%%%%%%%%%%%%%%%%%%%%%%%%%%%%%%%%%%%%%%%%%%%%%%
%%%%%%%%%%%%%%%%%%%%%%%%%%%%%%%%%%%%%%%%%%%%%%%%%%%%%%%%%%%%%%%%%%%%%%%%%
\begin{figure}
\includegraphics[width=0.67\columnwidth]{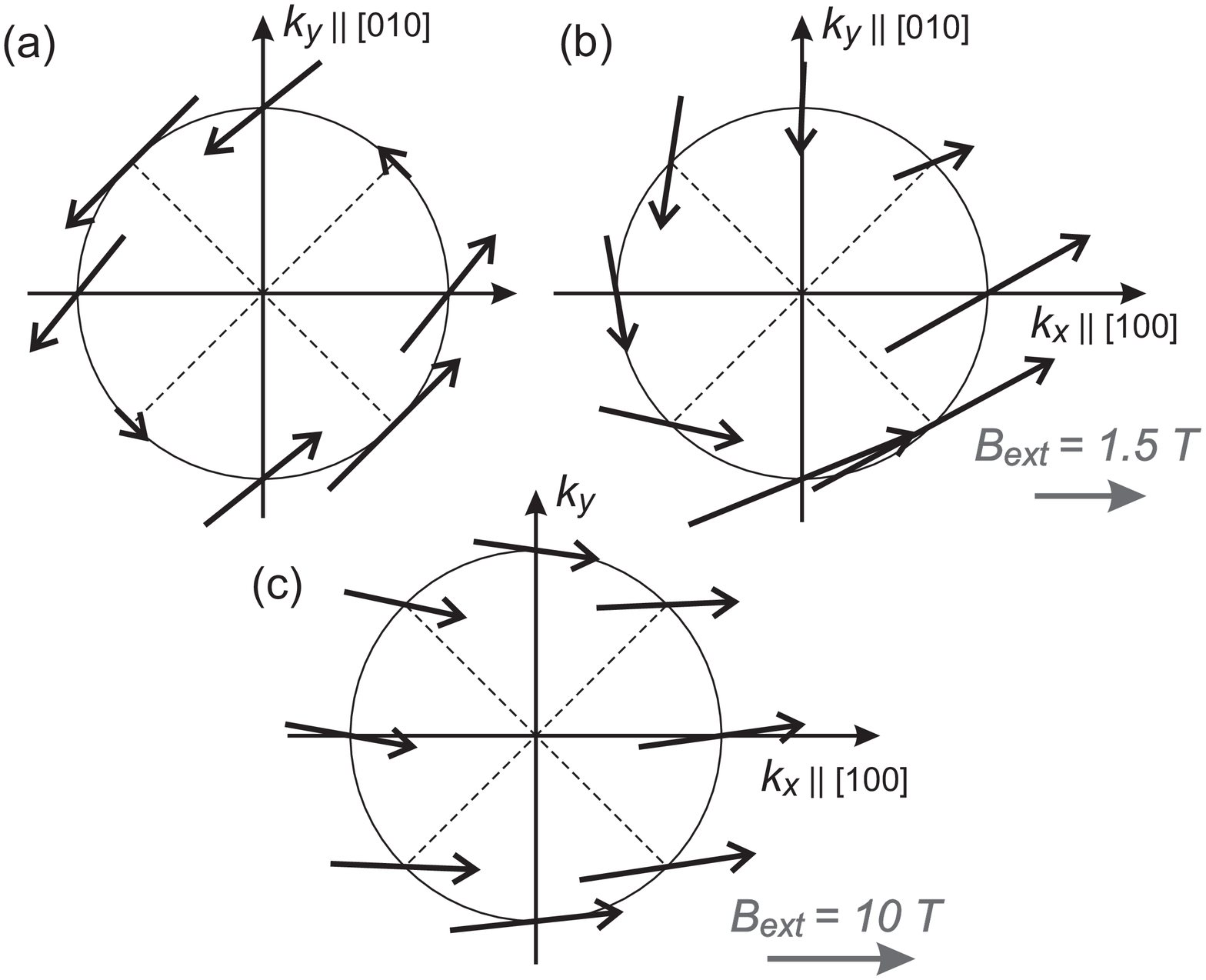}
\caption{(a) Schematic representation of the direction and
magnitude of the SOI fields. Since we assume all electrons
move with the same magnitude for $k$-vector $k_F$, we can
represent the motion of electrons in all directions as a
circle in the ($k_x$,$k_y$)-plane. The arrows that are
sketched at certain points on this Fermi circle, represent
the strength and direction of the SO field ${\rm {\bf
B}}_{SO}$ for that $k$-vector. (b) Magnitude and direction
of the total effective magnetic field when an external
field $B_{ext} = 1.5~{\rm T}$ is applied along the
[100]-direction. (c) Idem for an external field $B_{ext} =
10~{\rm T}$. In this plot the length of all arrows has been
scaled down by a factor 5 as compared to (a) and (b).}
\label{8Fig:SOfieldsIntroduction}
\end{figure}
%%%%%%%%%%%%%%%%%%%%%%%%%%%%%%%%%%%%%%%%%%%%%%%%%%%%%%%%%%%%%%%%%%%%%%%%%

A similar trend is observed when the degree of electron
confinement in a device structure is increased. Stronger
confinement gives more frequent scattering on the
boundaries of the system, and hence reduces the relaxation.
This has been recognized in the increase of $T_{1}$ on the
transition from 2D to 1D systems \cite{8Kiselev2000}. In
the limit quantum confinement in extremely small devices
(much smaller systems than we consider for the present
study), relaxation and dephasing due to SOI is then
strongly reduced, and other relaxation mechanisms can
become dominant. This applies for example to spin dephasing
in few-electron quantum dots, which can be mainly due to
interaction with nuclear spins
\cite{8Nazarov2000,8Nazarov2001,8Hanson2007}. Although more
frequent scattering due to stronger confinement thus seems
similar to reducing the mobility in bulk materials, the
results that we present here show that frequent scattering
due to confinement can also result in the opposite, namely
confinement-enhanced relaxation.

The key result of the present study is well presented by
the traces for spin relaxation time $T_1$ as a function of
the size $L$ of a square quantum dot in
Fig.~\ref{8Fig:Trelax_vs_length}a. The trace for the case
that an external magnetic field $B_{ext}=0~{\rm T}$ and
2DEG mobility $\mu =100~{\rm m^2/Vs}$, shows that $T_1$ is
constant for $L > 10~{\rm \mu m}$. Here $L$ is so large
that the electron ensemble behaves as in a free 2DEG. When
decreasing $L$ below $10~{\rm \mu m}$, $T_1$ increases
because spin relaxation in suppressed by more frequent
scattering on the edge of the system. The trace for
$B_{ext}=10~{\rm T}$ and 2DEG mobility $\mu =100~{\rm
m^2/Vs}$ (typical parameters for research on spin effects
in micronscale quantum dots \cite{8Koop2008} and wires
\cite{8Frolov2008}), however, shows radically different
behavior. Now $T_1$ first slowly decreases when decreasing
$L$ from the 2D regime (very large $L$), and shows a
pronounced dip for $L \approx 1~{\rm \mu m}$. Confinement
now strongly enhances relaxation, instead of the more
familiar result that confinement reduces relaxation.
Moreover, in the range with $1~{\rm \mu m} < L < 10~{\rm
\mu m}$, $T_{1}$ has a highly structured dependence on $L$.
Only when decreasing $L$ below $L \approx 1~{\rm \mu m}$,
$T_1$ shows again a strong and monotonic increase as for
the confinement-suppressed relaxation in $B_{ext}=0~{\rm
T}$.

We obtain these results with a numerical Monte Carlo
approach. This has the advantage that we can study
precessional relaxation for realistic conditions, where the
total magnetic field is the sum of several spin-orbit
contributions and an external field. In reality, samples
typically have both Rashba and Dresselhaus SOI that are
comparable in magnitude \cite{8Miller2003}, and in
experiments one often needs to apply strong external
magnetic fields for realizing spin transport in
non-magnetic semiconductors
\cite{8Potok2002,8Koop2008,8Frolov2008}. Earlier studies of
these relaxation phenomena were often restricted to more
tractable cases, as for example with only Rashba SOI
\cite{8Chang2004} (no Dresselhaus SOI), and no external
fields. Below, we will summarize our numerical method, and
then focus on studying the dependence of $T_{1}$ on the
degree of confinement in micronscale quantum dots. We will
also show that in regimes and for parameters that were
studied before, our simulations give the conventional
results.

%%%%%%%%%%%%%%%%%%%%%%%%%%%%%%%%%%%%%%%%%%%%%%%%%%%%%%%%%%%%%%%%%%%%%%%%%
%%% SPIN-ORBIT %%%%%%%%%%%%%%%%%%%%%%%%%%%%%%%%%%%%%%%%%%%%%%%%%%%%%%%%%%
%%%%%%%%%%%%%%%%%%%%%%%%%%%%%%%%%%%%%%%%%%%%%%%%%%%%%%%%%%%%%%%%%%%%%%%%%

We use a description where spin-orbit (SO) coupling acts as
a $k$-vector dependent effective magnetic field on the
electron spins. In a 2DEG it is dominated by two sources
\cite{8Miller2003}. The first arises due to the inversion
asymmetry in the potential profile of the heterostructure
and results in an effective Rashba magnetic field ${\rm
{\bf B}}_{R}$ \cite{8Rashba1984}. The second effect arises
due to the lack of inversion symmetry in the GaAs crystal
lattice, which is of the zinc-blende type, and yields for a
2DEG the linear and cubic Dresselhaus fields ${\rm {\bf
B}}_{D1}$ and ${\rm {\bf B}}_{D3}$
\cite{8Dresselhaus1955,8Silsbee2004}. The effective SO
field ${\rm {\bf B}}_{SO}$ in a 2DEG can thus be described
as the vector sum of these three components, given by
\cite{8Miller2003,8Silsbee2004}
   \begin{eqnarray} \label{8Eq:SORashba}
      {\rm {\bf B}}_{R} \quad = \quad C_{R} ({\rm {\bf \hat{x}}} k_y - {\rm {\bf \hat{y}}} k_x ),\\ \label{8Eq:SODresselhaus1}
      {\rm {\bf B}}_{D1} \quad = \quad C_{D1} (-{\rm {\bf \hat{x}}} k_x + {\rm {\bf \hat{y}}} k_y),\\ \label{8Eq:SODresselhaus3}
      {\rm {\bf B}}_{D3} \quad = \quad C_{D3} ({\rm {\bf \hat{x}}} k_x k_y^2 - {\rm {\bf \hat{y}}} k_y k_x^2),
   \end{eqnarray}
where $C_{R}$, $C_{D1}$ and $C_{D3}$ are the coupling
parameters, ${\rm {\bf \hat{x}}}$ is the unit vector in the
[100]-direction, and ${\rm {\bf \hat{y}}}$ in the
[010]-direction. Our results are calculated using the SO
parameters that were reported by Miller \textit{et al.}
\cite{8Miller2003}, $C_{R} = -1.96~\cdot~10^{-8}~{\rm Tm}$,
$C_{D1} = -1.57~\cdot~10^{-8}~{\rm Tm}$, and $C_{D3} =
-1.18~\cdot~10^{-24}~{\rm Tm^3}$.

The total SO field is then anisotropic in momentum space as
is shown in Fig.~\ref{8Fig:SOfieldsIntroduction}a. Each
time an electron scatters and its direction of motion
changes it will precess around a different axis set by
${\rm {\bf B}}_{SO}$. For our set of SO parameters these
fields ${\rm {\bf B}}_{SO}$ lie more or less parallel to
the [110]-direction for almost all $k$-directions.
Figures~\ref{8Fig:SOfieldsIntroduction}b,c show the effect
of adding an external magnetic field ${\rm {\bf B}}_{ext}$,
which is independent of momentum direction. We consider
here the situation that ${\rm {\bf B}}_{ext}
\parallel {\rm {\bf \hat{x}}}$. The total effective magnetic field ${\rm {\bf B}}_{tot}$
is then the vector sum of the SO fields and the external magnetic
field. When the magnitude $B_{ext}$ of ${\rm {\bf B}}_{ext}$ is
comparable to that of the SO fields, as shown in
Fig.~\ref{8Fig:SOfieldsIntroduction}b for $B_{ext} = 1.5~{\rm T}$,
the total effective magnetic fields are no longer mainly parallel to
the [110]-direction and there is larger spread in the directions of
the precession axes ${\rm {\bf B}}_{tot}$.
Figure~\ref{8Fig:SOfieldsIntroduction}c shows that for very large
external magnetic fields, shown here for $B_{ext} = 10~{\rm T}$, the
total effective magnetic fields ${\rm {\bf B}}_{tot}$ align with
${\rm {\bf B}}_{ext}$ along the [100]-direction. This again reduces
the spread in the direction of the precession axes ${\rm {\bf
B}}_{tot}$.

%%%%%%%%%%%%%%%%%%%%%%%%%%%%%%%%%%%%%%%%%%%%%%%%%%%%%%%%%%%%%%%%%%%%%%%%%
%%% MODEL %%%%%%%%%%%%%%%%%%%%%%%%%%%%%%%%%%%%%%%%%%%%%%%%%%%%%%%%%%%%%%%
%%%%%%%%%%%%%%%%%%%%%%%%%%%%%%%%%%%%%%%%%%%%%%%%%%%%%%%%%%%%%%%%%%%%%%%%%

In our numerical approach we use a classical description of
the electron motion, and a quantum mechanical description
of the dynamics of the electron spin. We thus assume that
electrons have at all times a well-defined $k$-vector, and
electrons move along classical trajectories with specular
scattering on the boundaries of the system, and scattering
in a random direction on static potential fluctuations due
to impurities. Electrons never escape from the system.
Although in micronscale quantum dots the motion of
electrons is confined, the mean level spacing is much
smaller than temperature, $\Delta_m~\ll~k_B T$, and this
allows for this semiclassical description
\cite{8Beenakker1997}. Further, we consider the case that
all the electrons that carry the spin orientation are near
the Fermi level. Thus, we assume that all electron always
move with the Fermi velocity ($k$-vectors with magnitude
$k_F$), independent of the momentum direction. This is a
valid approximation for $k_B T, \Delta E_{Z,SO}~\ll~E_F$
(with respect to the bottom of the conduction band), where
$\Delta E_{Z,SO}$ is the Zeeman splitting due to the SO
field alone. Obviously, the validity of our approach breaks
down in the limit of very small dots, where electrons are
highly localized due to quantum confinement. In practice,
this occurs for quantum dots with size $L$ below $\sim
400~{\rm nm}$, but our results for this regime always show
a very strong suppression of precessional relaxation, which
is the semiclassical equivalent for suppressed relaxation
for quantum confined electrons
\cite{8Nazarov2000,8Nazarov2001}.

Our simulation then works as follows. It starts at $t = 0$
with each electron at a random position in a square shaped
dot of size $L$, and with a $k$-vector in a random
direction. We always consider the case that at $t = 0$ the
spin state is prepared in the positive ${\rm {\bf
\hat{x}}}$-direction. For each electron, we follow its
state in time, and its momentum direction will change at
each scattering event. During each ballistic trajectory
between scatter events, the electron has a well-defined
$k$-vector, and we calculate the effective spin-orbit field
during this trajectory with
Eqs.~\ref{8Eq:SORashba}-\ref{8Eq:SODresselhaus3}. After
each scattering event the electron will thus precess around
a new effective magnetic field, and we follow the quantum
mechanical spin evolution in the total effective magnetic
field during each ballistic trajectory. We thus find the
spin state of each electron as a function of time.

The spin evolution is calculated for an ensemble of (at
least) $10^3$ electrons. This mimics the averaging over
many electrons in an electron transport experiment, since
large quantum dots behave in practice as a chaotic
ballistic cavity \cite{8Beenakker1997,8Koop2008}. In our
model the magnitude of the average spin orientation for the
ensemble decays to zero because each electron has its own
scattering trajectory, such that the relative difference
between the precessional dynamics of individual electrons
increases in time. We will always refer to this as spin
relaxation for the ensemble (with decay time $T_1$), rather
than dephasing, because we concentrate on the loss of
average spin orientation in the direction of the external
magnetic field that we apply. Note however, that in our
description each individual electron always keeps
precessing coherently in its particular spin-orbit field.
Consequently, the underlying mechanism is equivalent to
that of spin dephasing for an ensemble, and in particular
for our simulations with $B_{ext}=0~{\rm T}$ one could
argue that the loss of spin orientation should be named
dephasing.

The average spin orientation is calculated for the whole ensemble as
a function of time, independent of the position of the individual
electrons. In this study we concentrate on the average spin
polarization $\langle S_x \rangle$ in the ${\rm {\bf
\hat{x}}}$-direction. We find in all cases that we consider here
that the decay time for $\langle S_x \rangle$ equals that of
$\langle S \rangle$, where $\langle S \rangle = \sqrt{\langle S_x
\rangle^2 + \langle S_y \rangle^2 + \langle S_z \rangle^2}$, because
no significant polarization develops in the ${\rm { \bf \hat{y}}}$
or ${\rm { \bf \hat{z}}}$-direction. The relaxation time $T_1$ is
then defined as the time when $\langle S_x \rangle$ is reduced to
$1/e$ of its initial value at $t=0$. We used electron density $1.0
\cdot 10^{15}~{\rm m^{-2}}$ and mobility 100~${\rm m^2 / Vs}$,
unless stated otherwise. We neglect inelastic scattering mechanisms
and electron-electron interactions.

The momentum direction for an electron changes after
specular scattering on the boundary of the system. This
process we will refer to as edge scattering and the typical
length scale involved here is $L$, where the area of the
quantum dot system is $A = L^2$. A second effect causing a
change in momentum direction is scattering on static
fluctuations in the potential due to impurities. Here, a
scatter event changes the momentum into a random direction.
We incorporate this into our modeling as follows. When an
electron is moving ballistically through the system, the
probability that it did {\it not} scatter due to impurities
decreases as $e^{-t/\tau_{s}}$ (where $\tau_s$ is the
average impurity scatter time) and this probability is
reset to 1 after each impurity scatter event. We thus
define the mean free path $L_{mfp} = |v_F| \tau_{s}$ as the
length in between scatter events when only considering
impurity scattering.

%%%%%%%%%%%%%%%%%%%%%%%%%%%%%%%%%%%%%%%%%%%%%%%%%%%%%%%%%%%%%%%%%%%%%%%%%
%%% FIGURE FREE 2DEG %%%%%%%%%%%%%%%%%%%%%%%%%%%%%%%%%%%%%%%%%%%%%%%%%%%%
%%%%%%%%%%%%%%%%%%%%%%%%%%%%%%%%%%%%%%%%%%%%%%%%%%%%%%%%%%%%%%%%%%%%%%%%%
\begin{figure}[b]
\includegraphics[width=0.67\columnwidth]{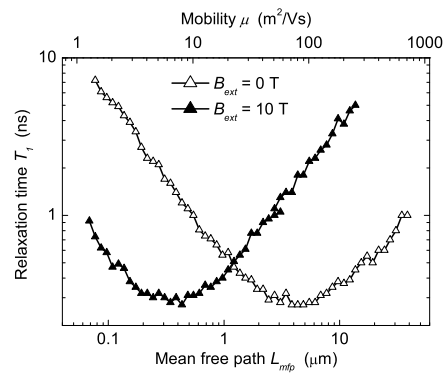}
\caption{Relaxation time $T_1$ as a function of the mean free path
$L_{mfp}$ for a free 2DEG. The top axis shows the corresponding
value for mobility $\mu$. Calculations for zero external field and
$B_{ext} = 10~{\rm T}$.} \label{8Fig:Trelax_vs_mobility}
\end{figure}
%%%%%%%%%%%%%%%%%%%%%%%%%%%%%%%%%%%%%%%%%%%%%%%%%%%%%%%%%%%%%%%%%%%%%%%%%

Another important length scale in our system is the
so-called precession length $\langle L_{pr} \rangle$. This
is defined as the length of the trajectory where a spin has
precessed over an angle $\pi$. In our system this length
scale is $k$-vector dependent due to the anisotropy of SO
fields. Therefore, we define $\langle L_{pr} \rangle$ as
the length of the trajectory for precession over an angle
$\pi$ around the average total effective magnetic field
$\langle |{\rm {\bf B}}_{tot}| \rangle$, where we average
over all $k$-directions to account for the anisotropy for
the SO contribution to the total field.

Evaluating the relative size of these three length scales,
$L$, $L_{mfp}$, and $\langle L_{pr} \rangle$, for a
specific system helps to understand many properties of the
relaxation time. Before discussing results for quantum
dots, it is instructive to discuss two regimes that occur
for a free 2DEG ($L$ very large). For such systems,
Fig.~\ref{8Fig:Trelax_vs_mobility} presents traces with
$T_1$ as a function of $L_{mfp}$. We first focus on the
case with $B_{ext} = 0~{\rm T}$, for which $\langle L_{pr}
\rangle = 8 ~ {\rm \mu m}$. If $L_{mfp} \ll \langle L_{pr}
\rangle$, the precession angle in between scatter events is
small. Consequently, the spin state only slowly diffuses
away from its initial direction in a random walk-like
process. Here scattering suppresses relaxation. This regime
is known as the motional narrowing regime
\cite{8Slichter1989}, and the relaxation time is inversely
proportional to the scatter time,
$T_1~\propto~\tau_{s}^{-1}$. When $L_{mfp} \gg \langle
L_{pr} \rangle$, the spins will coherently precess over at
least one full rotation between scatter events. Without
momentum scattering, the ensemble shows spin relaxation
since each electrons precesses in a different field, but
each electron will maintain its component in the direction
of its precession axis. Only when a scatter event occurs
this coherent precession is disturbed, such that further
relaxation for the ensemble can occur. In this regime
$T_1~\propto~\tau_{s}$. The crossover between these two
regimes (where $T_1$ shows a minimum) occurs for $L_{mfp}
\approx \langle L_{pr} \rangle$. Switching on an external
magnetic field of 10~T for this system, reduces the length
scale $\langle L_{pr} \rangle$ to $1.1 ~ {\rm \mu m}$. For
$T_1$ as a function of $L_{mfp}$, this only results in a
shift of the entire curve, with the minimum now occurring
at the new value where $L_{mfp} \approx \langle L_{pr}
\rangle$. We conclude here that our calculations for
Fig.~\ref{8Fig:Trelax_vs_mobility} reproduce the
conventional result \cite{8Fabian2007}.

%%%%%%%%%%%%%%%%%%%%%%%%%%%%%%%%%%%%%%%%%%%%%%%%%%%%%%%%%%%%%%%%%%%%%%%%%
%%% RESULTS QUASI 0D %%%%%%%%%%%%%%%%%%%%%%%%%%%%%%%%%%%%%%%%%%%%%%%%%%%%
%%%%%%%%%%%%%%%%%%%%%%%%%%%%%%%%%%%%%%%%%%%%%%%%%%%%%%%%%%%%%%%%%%%%%%%%%
%%%
%%% INTERNAL NOTE: NOT CLEAR YET WHY $L^{-2}$ AND
%%% NOT $L^{-1}$
%%%

We now turn to discussing results for quantum dots, for which $L$
can be smaller than $\langle L_{pr} \rangle$ and $L_{mfp}$.
Figure~\ref{8Fig:Trelax_vs_length}a shows simulations of the
relaxation time as a function of dot size $L$. We first discuss the
result for zero external magnetic field (which gives $\langle L_{pr}
\rangle = 8~{\rm \mu m}$) and mobility $\mu = 100~{\rm m^2/Vs}$
(corresponding to $L_{mfp} = 5~{\rm \mu m}$). For this system, a
decrease of $L$ in the regime with $L < \langle L_{pr} \rangle$
results in higher $T_1$ values. We will refer to this regime as the
quasi-0D regime. Here frequent scattering suppresses precession, as
for motional narrowing in the 2D regime. Here, we observe that $T_1
\propto L^{-2}$, a dependence on $L$ that was also found for studies
on the suppressed relaxation in long quasi-1D channels of width $L$
\cite{8Kiselev2000,8Malshukov2000}. For $L > \langle L_{pr} \rangle$
we do not observe any change in $T_1$. This will be denoted as the
2D regime, for which $L \gg \langle L_{pr} \rangle, L_{mfp}$.

For $B_{ext} = 10~{\rm T}$ (and again $\mu = 100~{\rm
m^2/Vs}$) the behavior is dramatically different (also
shown in Fig.~\ref{8Fig:Trelax_vs_length}a). When coming
from the 2D regime, there is no longer simply an increase
in $T_1$ when lowering $L$ towards the regime where $L \ll
\langle L_{pr} \rangle, L_{mfp}$. Instead, there is a
regime, here for $0.6~{\rm \mu m} < L < 10~ {\rm \mu m}$,
where $T_1$ is strongly suppressed. Moreover, the decrease
in $T_1$ when lowering $L$ from 10 $\mu m$ to 1 $\mu m$
shows a structured pattern. For this value of the external
magnetic field the precession length is reduced to $\langle
L_{pr} \rangle = 1.1 ~{\rm \mu m}$. The mean free path is
still $L_{mfp} = 5 ~{\rm \mu m}$, such that switching on a
strong field opens up a regime with $\langle L_{pr} \rangle
\lessapprox L < L_{mfp}$ in between the 2D and the quasi-0D
regimes. Notably, switching on a strong magnetic field for
a system in the 2D regime increases $T_1$ by about one
order of magnitude. However, switching on a field for a
system in the same material with $L \approx 1~{\rm \mu m}$
causes $T_1$ to go down more than 2 orders of magnitude.
Now confinement enhances relaxation.

For $B_{ext} = 0 ~ {\rm T}$ and $\mu = 100~{\rm m^2/Vs}$ we
do not see a dependence of $T_1$ on $L$ in the regime where
$L > \langle L_{pr} \rangle$, because for that system
$L_{mfp} \approx \langle L_{pr} \rangle$. However, also in
zero external magnetic field we can open up a regime where
$\langle L_{pr} \rangle \lessapprox L < L_{mfp}$ by
choosing a higher value for mobility. This is demonstrated
for $\mu = 1000~{\rm m^2/Vs}$ (which gives $L_{mfp} =
50~{\rm \mu m}$) in Fig.~\ref{8Fig:Trelax_vs_length}a. Now
for $7~{\rm \mu m} < L < 100~{\rm \mu m}$ the relaxation
time decreases when decreasing $L$, and again $T_1$ shows a
structured pattern (\textit{i.e.} the structure on this
trace here is not noise from averaging over a finite
ensemble).

The reduction in $T_1$ due to stronger confinement is thus
a general effect and appears whenever $\langle L_{pr}
\rangle \lessapprox L < L_{mfp}$. These are typical
conditions for micronscale quantum dots when large external
magnetic fields are applied. Notably, the values that we
obtain here for $T_1$ are very close to the values that we
recently observed in spin accumulation experiments in
micronscale quantum dots \cite{8Koop2008}. We found $T_1
\approx 300~{\rm ps}$ for a quantum dot with $L \approx 1.1
~{\rm \mu m}$ and $B_{ext} = 8.5 ~{\rm T}$. This indicates
that our simulations generate realistic numbers.

%%%%%%%%%%%%%%%%%%%%%%%%%%%%%%%%%%%%%%%%%%%%%%%%%%%%%%%%%%%%%%%%%%%%%%%%%
%%% RESULTS ONLY RASHBA %%%%%%%%%%%%%%%%%%%%%%%%%%%%%%%%%%%%%%%%%%%%%%%%%
%%%%%%%%%%%%%%%%%%%%%%%%%%%%%%%%%%%%%%%%%%%%%%%%%%%%%%%%%%%%%%%%%%%%%%%%%

We will now analyze the relaxation mechanism for this
confinement-enhanced relaxation. To study why there is
strong dip and structure in the dependence of $T_1$ on $L$,
we choose a simplified model system where we only consider
the Rashba SOI and an external magnetic field $B_{ext} =
10~{\rm T}$. Both Dresselhaus SOI contributions have been
set to zero. Using only Rashba has the advantage that the
magnitude for the SO fields is identical for all
$k$-vectors. The structure in $T_1$ appears more regular
here (see Fig.~\ref{8Fig:Trelax_vs_length}b). This proves
that the effects that we present here do not only occur for
very particular SO parameters.

We have repeated this calculation where we programmed non-specular
reflections on the walls of the quantum dot, such that after hitting
a side of the quantum dot the electron is reflected with random
angle back into the quantum dot (open symbols in
Fig.~\ref{8Fig:Trelax_vs_length}b). The structure on $T_1$ still
appears, which confirms that self-repeating patterns are not the
origin of this effect. We observe, however, that the amplitude of
the structure on $T_1$ is reduced for this setting. This is caused
by an increased variation in the length of trajectories in between
scatter events for non-specular reflections. We checked this by
making histograms of the trajectory lengths for specular and
non-specular reflections (not shown).

For both traces in Fig.~\ref{8Fig:Trelax_vs_length}b, we find that
the minimums in $T_1$ appear at odd multiples of the average
precession length $\langle L_{pr} \rangle$, and local maximums occur
at even multiples of $\langle L_{pr} \rangle$. This means that for
systems with a size equal to an odd multiple of $\langle L_{pr}
\rangle$ the electrons scatter (on average) after precessing (again,
on average) over an angle of exactly $\pi~({\rm mod}~2 \pi)$ between
scatter events, furthest away from their original state. This causes
fast relaxation. For systems with a size equal to an even multiple
of $\langle L_{pr} \rangle$, the electrons scatter on average after
precessing $2 \pi~({\rm mod}~2 \pi)$, so when they are back in their
original state. Then relaxation is slower as compared to the local
minimums. However, note that for these local maximums there is
overall still a reduction of $T_1$ due to confinement as compared to
free 2DEG: in this regime, more frequent scattering on the edge of
the system always enhances relaxation. The character of the
relaxation mechanism itself is thus similar to the regime with
$\langle L_{pr} \rangle \ll L_{mfp}$ for 2D systems, where $T_1
\propto \tau_s$. The additional feature here is the structure on
$T_1$ as a function of $L$, which signals that the overall
relaxation mechanism is either somewhat resonantly enhanced or
suppressed when the time-of-flight across the dot matches even or
odd multiples of the spin precession time for an angle $\pi$.
Notably, the results for free 2DEG in the regime with $T_1 \propto
\tau_s$ (Fig.~\ref{8Fig:Trelax_vs_mobility}) do not show structure
on $T_1$ because there is a larger spread in the scatter times
$\tau_s$. We checked that when we program that impurity scattering
in a random direction always occurs after a fixed time $\tau_s$ (no
spread), we also observe structure on $T_1$ as a function of
$L_{mfp}$ for a free 2DEG (not shown).

The most extreme suppression of $T_1$ due to confinement is in
ballistic quantum dots ($L \ll L_{mfp}$) when the size of the system
$L=\langle L_{pr} \rangle$. Figure~\ref{8Fig:SOfieldsIntroduction}c
indicates that this is a counter-intuitive result when this
condition is met in strong external fields. The various precession
axes get more and more aligned when the external field is increased
to 10~T in ${\rm {\bf \hat{x}}}$-direction, while the spins are
prepared in this direction. Nevertheless, the lowest $T_1$ value
that occurs for the various traces in
Fig.~\ref{8Fig:Trelax_vs_length}a is for a quantum dot system of $L
= 1.1 ~ {\rm \mu m}$ in a field of $B_{ext}=10~{\rm T}$. Due to the
initial spin state in the ${\rm {\bf \hat{x}}}$-direction, spins are
initially precessing with relatively small cone angles around these
effective magnetic fields, and all electrons will maintain a large
component in the ${\rm {\bf \hat{x}}}$-direction. This results only
in a small reduction of $\langle S_x \rangle$. Further relaxation
for the ensemble only progresses when spins hop onto wider
precession cone angles, which only occurs at a scatter event. Thus,
more scattering leads to more rapid relaxation, in particular when
$L$ is an odd multiple of $\langle L_{pr} \rangle$, and most rapidly
when $L = \langle L_{pr} \rangle$. The reason that this results in a
very fast relaxation mechanism for small systems in strong magnetic
fields is that the precession and scatter times that underlie this
mechanism are then very short.

It is interesting to note that a similar conclusion was
reached with a very different approach in work that studied
how conductance fluctuations of large quantum dots are
influenced by spin-orbit effects and strong in-plane
magnetic fields. Here, it was found that applying a strong
in-plane magnetic field can enhance the suppression of
conductance fluctuations by spin-orbit effects
\cite{8Folk2001}. Theoretical work on this phenomenon
reached the conclusion that this effect is strongest in
ballistic dots, where the precession time is on the order
of the time of flight across the dot \cite{8Halperin2001}.

%%%%%%%%%%%%%%%%%%%%%%%%%%%%%%%%%%%%%%%%%%%%%%%%%%%%%%%%%%%%%%%%%%%%%%%%%
%%% CONCLUSIONS %%%%%%%%%%%%%%%%%%%%%%%%%%%%%%%%%%%%%%%%%%%%%%%%%%%%%%%%%
%%%%%%%%%%%%%%%%%%%%%%%%%%%%%%%%%%%%%%%%%%%%%%%%%%%%%%%%%%%%%%%%%%%%%%%%%

In conclusion, we have shown that the relaxation time $T_1$
for a spin population in a certain 2DEG material can be
strongly decreased when bringing the size of the system
from free 2DEG down to micronscale quantum dots. The
strongest suppression is found for ballistic systems of a
size $L$ that equals the precession length $\langle L_{pr}
\rangle$. Here, frequent scattering on the edge of the dot
rapidly drives precession onto wider and wider cone angles,
and this effect is resonantly enhanced in all systems where
$L$ equals an odd multiple of $\langle L_{pr} \rangle$. We
believe our results are very useful for comparison to
experimental results on this type of systems, since we can
use realistic device and SO parameters. Furthermore, the
$T_1$ values that we calculate match very well with our
recent experimental results on micronscale quantum dots
\cite{8Koop2008}.

We thank M. J. van Veenhuizen, A. I. Lerescu, J. Liu and T. Last for
useful discussions. This work was supported by the Dutch Foundation
for Fundamental Research on Matter (FOM) and the Netherlands
Organization for Scientific Research (NWO). During the preparation
of this manuscript we became aware of similar work underway by
S.~L\"{u}scher \textit{et al.} \cite{8Folk2008}.

%%%%%%%%%%%%%%%%%%%%%%%%%%%%%%%%%%%%%%%%%%%%%%%%%%%%%%%%%%%%%%%%%%%%%%%%%
%%% REFERENCES %%%%%%%%%%%%%%%%%%%%%%%%%%%%%%%%%%%%%%%%%%%%%%%%%%%%%%%%%%
%%%%%%%%%%%%%%%%%%%%%%%%%%%%%%%%%%%%%%%%%%%%%%%%%%%%%%%%%%%%%%%%%%%%%%%%%


\begin{references}

%1
\bibitem{8Silsbee2004}      For a recent review see R. H. Silsbee, J. Phys.: Cond. Mat. \textbf{16}, R179 (2004).
%2
\bibitem{8Fabian2007}       For a recent review see J. Fabian, A. Matos-Abiague, C. Ertler, P. Stano, I. Zutic, Acta Physica Slovaca {\bf 57}, 565 (2007); arXiv:0711.1461 (2007).
%3
\bibitem{8Datta1990}        S. Datta and B. Das, Appl. Phys. Lett {\bf 56}, 665 (1990).
%4
\bibitem{8Zutic2004}        I. \^{Z}uti\'{c}, J. Fabian, and S. Das Sarma, Rev. Mod. Phys. \textbf{76}, 323 (2004).
%5
\bibitem{8Awschalom2007}    D. D. Awschalom and M. E. Flatt\'{e}, Nature Phys. \textbf{3}, 153 (2007).
%6
\bibitem{8Dyakonov1971}     M. I. D'yakonov and V. I. Perel', Sov. Phys. JETP {\bf 33}, 1053 (1971); Sov. Phys. Solid State {\bf 13}, 3023 (1972).
%7
\bibitem{8Dyakonov1986}     M. I. D'yakonov and V. Y. Kachorovskii, Sov. Phys. Semicond {\bf 20}, 110 (1986).
%8
\bibitem{8Kiselev2000}      A. A. Kiselev and K. W. Kim, Phys. Rev. B {\bf 61}, 13115 (2000).
%9
\bibitem{8Hanson2007}       R. Hanson \textit{et al.}, Rev. Mod. Phys. {\bf 79}, 1217 (2007).
%10
\bibitem{8Nazarov2000}      A. V. Khaetskii and Y. V. Nazarov, Phys. Rev. B {\bf 61}, 12639 (2000).
%11
\bibitem{8Nazarov2001}      A. V. Khaetskii and Y. V. Nazarov, Phys. Rev. B {\bf 64}, 125316 (2001).
%12
\bibitem{8Koop2008}         E. J. Koop \textit{et al.}, arXiv:0801.2699 (2008).
%13
\bibitem{8Frolov2008}       S. M. Frolov \textit{et al.}, arXiv:0801.4021 (2008).
%14
\bibitem{8Miller2003}       J. B. Miller \textit{et al.}, Phys. Rev. Lett. {\bf 90}, 076807 (2003). %D.M. Zumb\"{u}hl, C.M. Marcus, Y.B. Lyanda-Geller, D. Goldhaber-Gordon, K. Chapman and A.C. Gossard
%15
\bibitem{8Potok2002}        R. M. Potok \textit{et al.}, Phys. Rev. Lett. \textbf{89}, 266602 (2002).
%16
\bibitem{8Chang2004}        Cheng-Hung Chang, A. G. Mal'schukov, and K. A. Chao, Phys. Rev. B {\bf 70}, 245309 (2004).
%17
\bibitem{8Rashba1984}       Y. A. Bychkov and E. I. Rashba, Sov. Phys. JETP, {\bf 39}, 78 (1984).
%18
\bibitem{8Dresselhaus1955}  G. Dresselhaus, Phys. Rev. {\bf 100}, 580 (1955).
%19
\bibitem{8Beenakker1997}    C. W. J. Beenakker, Rev. Mod. Phys. \textbf{69}, 731 (1997).
%20
\bibitem{8Slichter1989}     C. P. Slichter, {\it Principles of Magnetic Resonance}, Springer series in Solid-State Sciences Vol. 1, 3rd ed. (Springer-Verlag, Berlin, 1989), p. 213.
%21
\bibitem{8Malshukov2000}    A. G. Mal'shukov and K. A. Chao, Phys Rev. B \textbf{61}, R2413 (2000).
%22
\bibitem{8Folk2001}         J. A. Folk, S. R. Patel, K. M. Birnbaum, C. M. Marcus, C. I. Duru\"{o}z, and J. S. Harris, Phys. Rev. Lett. \textbf{86}, 2102 (2001).
%23
\bibitem{8Halperin2001}     B. I. Halperin, A. Stern, Y. Oreg, J. N. H. J. Cremers, J. A. Folk, and C. M. Marcus, Phys. Rev. Lett. \textbf{86}, 2106 (2001).
%24
\bibitem{8Folk2008}         S. L\"{u}scher, S. M. Frolov, and J. A. Folk (unpublished).



\end{references}
\end{document}